\documentclass[structabstract]{aa}  

\usepackage[tbtags,fleqn]{amsmath}
\usepackage{amsfonts}
\usepackage{graphicx}
\usepackage{natbib}

\begin{document}

   \title{Orion Revisited}

   \subtitle{I. The massive cluster in front of the Orion Nebula Cluster\thanks{Based on observations collected at the German-Spanish Astronomical Center, Calar Alto, jointly operated by the Max-Planck-Institut f\"ur Astronomie Heidelberg and the Instituto de Astrof\'{\i}sica de Andaluc\'{\i}a (CSIC).}} 
   \titlerunning{Orion Nebula Cluster Revisited}
   \author{Jo\~ao Alves
          \inst{1}
          \and
          Herv\'e Bouy\inst{2}
          }

   \institute{Department of Astrophysics, University of Vienna,
              T\"urkenschanzstrasse 17, 1180 Vienna, Austria\\
              \email{joao.alves@univie.ac.at}
         \and
              Centro de Astrobiolog\'{i}a, INTA-CSIC, PO Box 78, 28691 Villanueva de la Ca\~nada, Madrid, Spain \\
             \email{hbouy@cab.inta-csic.es}
             }

   \date{Received 27 July 2012; Accepted 21 September 2012} 


  \abstract
   {} 
  {The aim of this work is to characterize the stellar population between Earth and the Orion A molecular cloud where the well known star formation benchmark Orion Nebula Cluster (ONC) is embedded.}
  {We use the denser regions the Orion A cloud to block optical background light, effectively isolating the stellar population in front of it. We then use a multi-wavelength observational approach to characterize the cloud's foreground stellar population.}
  {We find that there is a rich stellar population in front of the Orion A cloud, from B-stars to M-stars, with a distinct 1) spatial distribution, 2) luminosity function, and 3) velocity dispersion from the reddened population inside the Orion A cloud. The spatial distribution of this population peaks strongly around NGC 1980 (iota Ori) and is, in all likelihood, the extended stellar content of this poorly studied cluster. We infer an age of $\sim 4-5$ Myr for NGC 1980 and estimate a cluster population of the order of 2000 stars, which makes it one of the most massive clusters in the entire Orion complex.   This newly found population overlaps significantly with what is currently assumed to be the ONC and the L1641N populations, and can make up for more than 10-20\% of what is currently taken in the literature as the ONC population (30-60\% if the Trapezium cluster is removed from consideration).  What is currently taken in the literature as the ONC is then a mix of several intrinsically different populations, namely: 1) the youngest population, including the Trapezium cluster and ongoing star formation in the dense gas inside the nebula, 2) the foreground population, dominated by the NGC 1980 cluster, and 3) the poorly constrained population of foreground and background Galactic field stars. 
 }
 { Our results support a scenario where the ONC and L1641N are not directly associated with NGC 1980, i.e., they are not the same population emerging from its parental cloud, but are instead distinct overlapping populations. The nearest massive star formation region and the template for massive star and cluster formation models is then substantially contaminated by the foreground stellar population of the massive NGC 1980 cluster, formed about 4--5 Myr ago in a different, but perhaps related, event in the larger Orion star formation complex. This result calls for a revision of most of the observables in the benchmark ONC region (e.g., ages, age spread, cluster size, mass function, disk frequency, etc.).  }


\maketitle

%

\section{Introduction}
\label{sec:introduction}

The Orion Nebula is one of the most studied objects in the sky, with observational records dating about 400 years coinciding with the early developments of the telescope \citep{Muench2008}. It is an object of critical importance for astrophysics as it contains the nearest (400 pc) massive star formation region to Earth, the Orion Nebula Cluster (ONC) \citep[e.g.][]{1965ApJ...142..964J,1972ApJ...175...89W}, which is the benchmark region for massive star and cluster formation studies. Recent distance estimates to the Orion Nebula using parallax put this object at about 400 pc from Earth (389$^{+24}_{-21}$ pc \citep{2007ApJ...667.1161S}, 414$\pm$7 pc \citep{Menten2007}, 437$\pm$19 pc \citep{2007PASJ...59..897H}, and 419$\pm$6 pc \citep{2008PASJ...60..991K}). Some of the most basic observables of the star formation process, like, 1) star formation rates \citep{Lada1995,Lada2010}, 2) star formation history \citep{Hillenbrand1997}, 3) age spreads \citep{Jeffries11,Reggiani11}, 4) the initial mass function to the substellar regime \citep{2000ApJ...540..236H,2002ApJ...573..366M,2012ApJ...748...14D,2012ApJ...752...59H}, 5) the fraction, size distribution, and lifetime of circumstellar disks \citep{Hillenbrand1998b,Lada2000,Muench2001,Vicente2005}, 6) their interplay with massive stars \citep{Odell1993}, binarity \citep{1998ApJ...500..825P,2006A&A...458..461K}, rotation \citep{2002A&A...396..513H}, magnetic fields \citep{2003ApJ...584..911F}, and 7) young cluster dynamics \citep{Hillenbrand1998,2008ApJ...676.1109F,2009ApJ...697.1103T}, have all been derived from this benchmark region \citep[see the meticulous reviews of][]{Bally2008,Muench2008,ODell2008}. Naturally, the ONC is also the benchmark region for theoretical and numerical models of massive and clustered star formation \citep{Palla1999,Klessen2000,Clarke2000,Bonnell2001,Bate2003,Tan2006,Huff2006,Krumholz2011,Allison2011}

\begin{figure*}[!tbp]
  \begin{center}
    \includegraphics[width=7in]{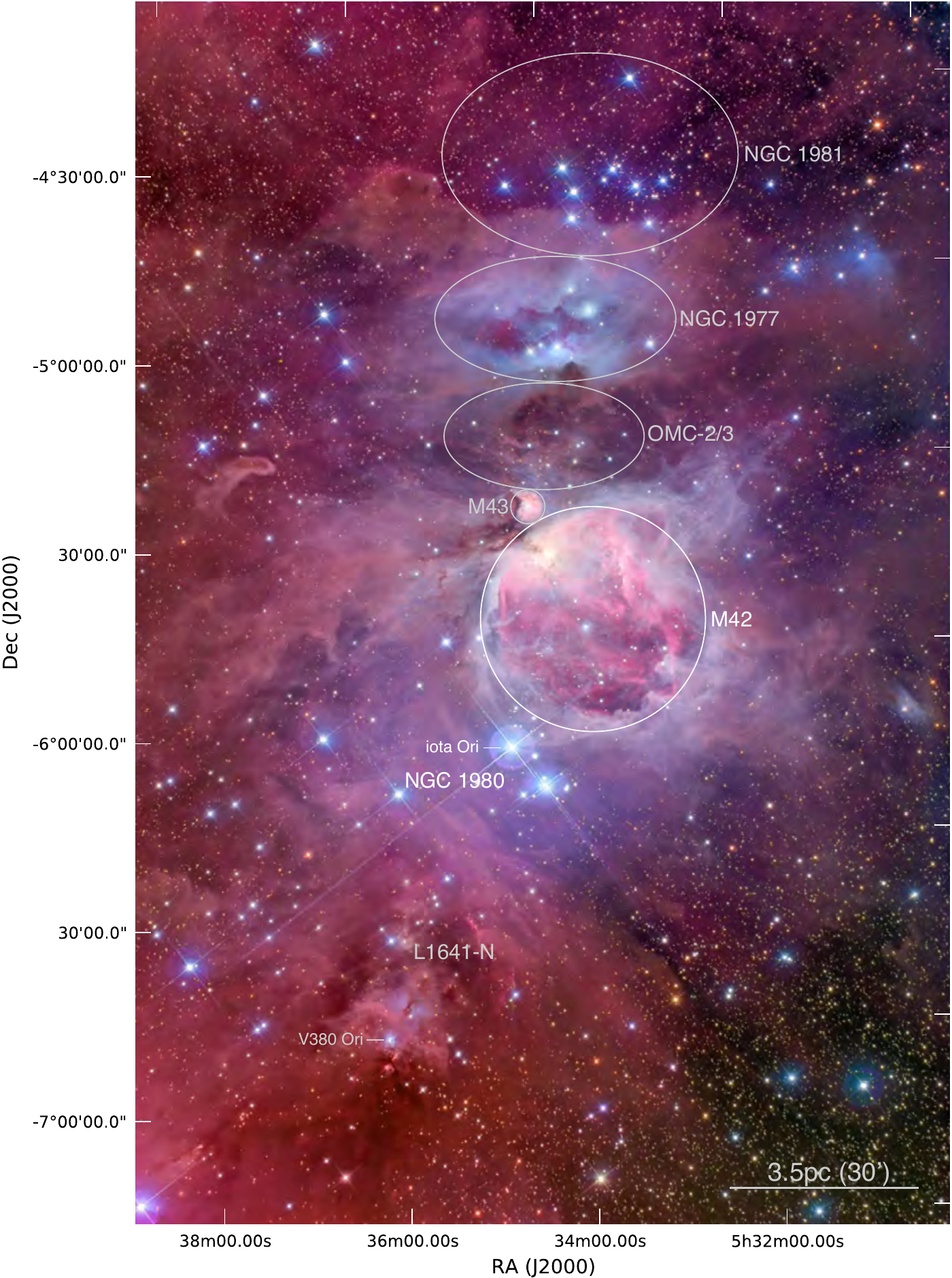}
    \caption{Optical image of the North end of the Orion A molecular cloud, including the relatively more evolved populations of NGC 1981, NGC 1977 and NGC 1980 (Orion OB 1c subgroup) and the Orion Nebula Cluster (Orion OB 1d subgroup), projected against the Orion Nebula (M42). This image illustrates well the complicated distribution of young stars in the vicinity of the ONC, with scattered groups of more evolved blue massive stars projected against partially embedded groups of younger stars (M43, ONC, OMC-2/3, L1641N). Image courtesy of Jon Christensen (christensenastroimages.com)
}
    \label{fig:1}
  \end{center}
\end{figure*}

\begin{figure}[!tbp]
  \begin{center}
    \includegraphics[width=\hsize,angle=0]{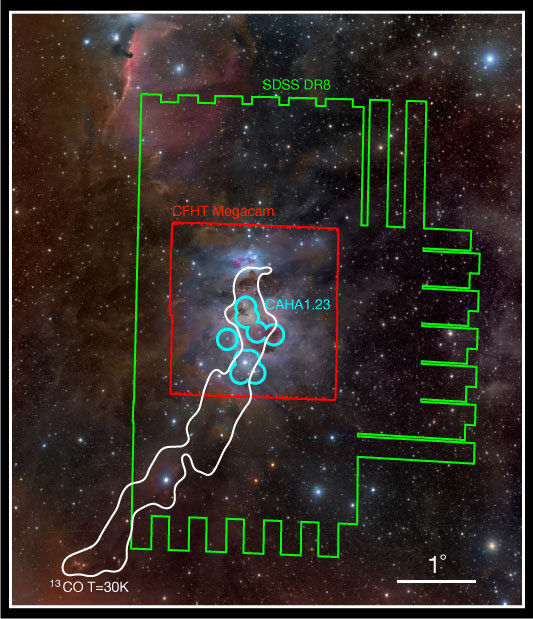}
    \caption{Coverage of the optical datasets used in this study. The SDSS images used in this study are represented in green, the {\it CFHT/Megacam} images in red, and the Calar Alto 1.23m CCD observations are represented in light blue. The contour corresponding to an integrated intensity of $^{13}$CO of 30~K km/s is represented in white. North is up and east is left. The angular scale is indicated in the lower left. Background photograph courtesy of Rogelio Bernal Andreo (DeepSkyColors.com) }
    \label{fig:coverage}
  \end{center}
\end{figure}

It is quite remarkable that within only 1.5$^{\circ}$ of the ONC there are several contiguous, and likely overlapping, groups of young stars (see Figure~\ref{fig:1}), although few studies have tackled the entire region as a whole.  Unfortunately the three-dimensional arrangement of star formation regions, in particular massive ones, is far from simple and is essentially unknown given the current distance accuracy to even the nearest star formation regions. It is clear however that the Orion Nebula cluster is partially embedded in its parental Orion A molecular cloud which in turn is inside the large $\sim 200$ pc Orion star formation complex, where groups of young stars with ages from a few to about 10 Myr are seen \citep{Brown1994,Briceno2007}. It has long been suspected that a more evolved group \citep[subgroup Ori OB 1c, including NGC 1981 and NGC 1980,][]{1964ARA&A...2..213B,1978ApJS...36..497W} is in the foreground of the molecular cloud where the younger ONC population (subgroup Ori OB 1d) is still partially embedded \citep[see][for a large scale analysis of the possible interplay between these two subgroups]{Gomez1998}.

There are two different views on the stellar population inside the Orion Nebula. The first suggests that the core of the ONC, the Trapezium cluster, is a distinct entity from the rest of the stellar population in the nebula, while the second, more prevalent, suggests the Trapezium is instead the core region of a larger cluster emerging from the Orion Nebula.  \citet{1986ApJ...307..609H} performed one of the first CCD observations of an area centered on the Trapezium cluster (covering $\sim9.2^{\prime \,2}$), and from the exceptional high stellar density found they argued that the Trapezium cluster was a distinct entity from the surrounding stellar population, including the stellar population inside the Orion Nebula. An opposite view was proposed by \citet{Hillenbrand1998} who compared optical and near-infrared surveys of the ONC with virial equilibrium cluster models to argue that the entire ONC is likely a single young stellar population.

Confirming which view is correct is critical because they imply different formation scenarios for the ONC, and assuming the ONC is typical, different scenarios for the formation of stellar clusters in general. While the first view implies the bursty formation of the bulk of the stars in a relatively small volume of the cloud, the second, by assuming a more extended cluster, calls necessarily for a longer and more continuos process, allowing for measurable age spreads in the young population, and for substancial fractions of young stellar objects (YSOs) at all evolutionary phases, from Class 0 to Class III. Observationally, the first view argues that the Orion OB 1c subgroup is a distinct star formation event from the 1d subgroup while the second and more prevalent view argues that the two subgroups are the same population, i.e., the Ori OB 1c subgroup is simply the more evolved stellar population emerging from the cloud where group 1d still resides.  

If the first view prevails, i.e., if the Trapezium cluster and ongoing star formation in the dense gas in its surroundings represent a distinct population from the rest of the stars in the larger ONC region, then what is normally taken in the literature as the ONC is likely to be a  superposition of different stellar populations. If this is the case, then the  basic star formation observables currently accepted for this benchmark region (e.g., ages, age spread, cluster size, mass function, disk frequency, etc.) could be compromised.

In this paper we address this important issue by attempting to characterize the stellar populations between Earth and the Orion Nebula. Our approach consists of using the Orion A cloud to block optical background light, effectively isolating the stellar population in front of it. We then use a multi-wavelength observational approach to characterize the cloud's and nebula's foreground population.  We find that there are two well defined, distinct, and unfortunately overlapping stellar populations: 1) a foreground, ``zero'' extinction population dominated by the poorly studied but massive NGC 1980 cluster, and 2) the reddened population associated with the Trapezium cluster and L1641N star forming regions, supporting the first view on the structure of the ONC as described above. This result calls for a revision of most of the star formation observables for this fiducial object.
  
This paper is structured as follows. In Sect, 2 we describe the observational data acquired for this project as well as the archival data used. In Sect. 3 we present the results of our approach, namely the identification of the two foreground populations and its characterization. We present a general discussion on the importance of the result found in Sect. 4 and summarize the main results of the paper in Sect. 5.

\begin{figure*}[!tbp]
  \begin{center}
    \includegraphics[width=\hsize,angle=0]{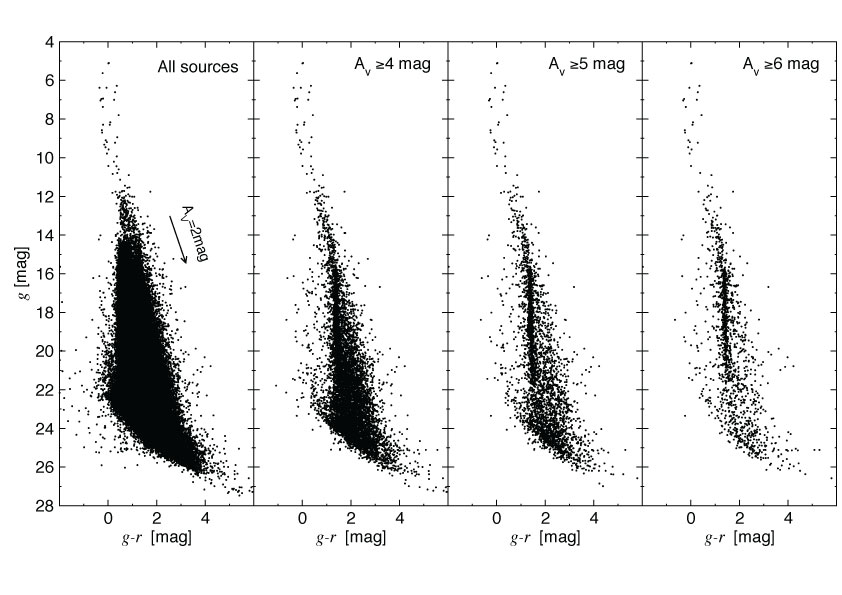}
    \caption{$g$ vs. $g-r$ color-magnitude diagram for the entire survey in regions of increasing total line-of-sight extinction.}
    \label{fig:grcmd}
  \end{center}
\end{figure*}

\section{Data}
\label{sec:data}

To characterize the foreground population to the Orion A molecular cloud we will make use of existing surveys together with raw data from Canada-France-Hawaii Telescope (CFHT), Calar Alto Observatory (CAHA 1.23m), and the Spitzer satellite, that were processed and analyzed for the purpose of this investigation.  

\subsection{Catalogues}
\label{sec:catalogues}
We retrieved the astrometry and photometry for all sources within a box of 5\degr$\times$15\degr\, centered around RA=85.7 and Dec=-4\degr (J2000) in the {\it Sloan Digital Sky Survey III}, the {\it Wide-field Infrared Survey Explorer} \citep[WISE ][]{2012yCat.2311....0C},  the Third XMM-Newton serendipitous source \citep{2009A&A...493..339W} 
and the 2MASS catalogues \citep{2006AJ....131.1163S}. Table~\ref{table_obs} gives an overview of the properties of these catalogues.

\begin{table}
\caption{Catalogues and observations used in this study \label{table_obs}}
\begin{tabular}{lc}\hline\hline
 Instrument     &  Band /   \\
                & Channel   \\
\hline
XMM-Newton/EPIC & 0.1--10~$keV$ \\
SDSS            & $u$,$g$,$r$,$i$,$z$   \\
CFHT/MegaCam    & $u$,$g$,$r$     \\
2MASS           & $J$,$H$,$Ks$ \\
WISE            & 3.3,4.6,12,22~$\mu$m  \\
Spitzer/IRAC    & 3.6,4.5,5.8,8.0~$\mu$m \\
Spitzer/MIPS    & 24~$\mu$m   \\
Calar Alto 1.23m & $u$,$g$,$r$ \\ 
\hline
\end{tabular}
\end{table}

\subsection{CFHT/Megacam}
\label{sec:cfhtmegacam}
A mosaic of 2$\times$2 pointings covering 2\degr$\times$2\degr\, centered on the Orion Nebula Cluster (ONC) was observed with {\it CFHT/Megacam} \citep{2003SPIE.4841...72B} with the Sloan $ugr$ filters on 2005 February 14 (P.I. Cuillandre). Figure~\ref{fig:coverage} gives an overview of the area covered by these observations. The conditions were photometric, as described in the \emph{Skyprobe} database \citep{2004ASSL..300..287C}. Seeing was variable, oscillating between 1--2\arcsec\, as measured in the images. A total of 5 exposures of 150~s ($u$-band), 40~s ($g$-band), and 40~s ($r$-band) each were obtained at each of the 4 positions. The observations were made in dither mode, with a jitter width of a few arcminutes at each position. This allows filling the CCD-to-CCD and position gaps and correcting for deviant pixels and cosmic ray events. The images were processed using the recommended \emph{Elixir} reduction package \citep{2004PASP..116..449M}. Aperture photometry was then extracted using {\it SExtractor} \citep{1996A&AS..117..393B} and the photometric zero-points in the SDSS system were derived by cross-matching with the SDSS catalogue. The CFHT/Megacam observations complement the SDSS data in one critical aspect: they provide data for regions around bright stars and nebulae, in particular the Orion Nebula region that is missing in the SDSS data.

\subsection{Calar Alto/1.23m CCD Camera}
\label{sec:caha123}
Selected pointings of the ONC (see Table~\ref{table_obs_caha} and Fig.~\ref{fig:coverage}) were observed on 2011 December 15  with the Calar Alto CCD camera mounted on the 1.23m telescope (hereafter CAHA123). The CCD camera is a 2k$\times$2k optical imager with a 17\arcmin\, field-of-view. The Sloan filters available at Calar Alto vignet the field and reduce it to a circular 11\arcmin\, diameter field-of-view. These observations are meant to complement the CFHT and SDSS observations below their saturation limits (at $ugr\approx$12~mag), and in the vicinity of bright saturated stars. Short exposures of 0.1 and 5.0~s were obtained in the Sloan $gr$ filters, and of 0.1 and 10~s in the Sloan $u$ filter. The telescope was slightly defocused to avoid saturation of the brightest stars. Three standard fields \citep[SA~97, SA~92 and BD+21D0607,][]{2002AJ....123.2121S} were observed during the course of the night to derive accurate zero-points. Each pointing was observed with a small dithering of a couple of arcminutes in order to correct for deviant pixels and cosmic ray events. The images were pre-processed (bias subtraction and flat-field correction) using standard procedure with the \textit{Eclipse} reduction package \citep{1997Msngr..87...19D}. The astrometric registration and stacking were then performed using the \textit{AstrOmatic} software suite \citep[][]{2010jena.confE.227B}. Aperture photometry was finally extracted using \textit{SExtractor} and the photometric zero-points in the SDSS system were derived by cross-matching with the SDSS and {\it Megacam} catalogues. The night was clear but not photometric. We observe a dispersion in the zero-point measurements through the night of 0.06~mag in $u$ and $g$, and 0.16~mag in $r$, which we add quadratically to the photometric measurement uncertainties. 

\begin{table}
\caption{CAHA 1.23m CCD observations\label{table_obs_caha}}
\begin{tabular}{lcc}\hline\hline
 Field          &  RA (J2000)        & Dec (J2000)             \\
\hline
Trapezium       & 05:35:19.341  &  $-$05:23:30.35 \\
Field 1         & 05:35:24.651  &  $-$05:55:06.69 \\
Field 2         & 05:34:56.819  &  $-$05:59:59.55  \\
Field 3         & 05:35:25.044  &  $-$05:59:15.44 \\
Field 4         & 05:34:52.120  &  $-$05:34:04.66 \\
Field 5         & 05:33:59.980  &  $-$05:35:40.17 \\
Field 6         & 05:36:09.483  &  $-$05:38:17.68 \\
Field 7         & 05:37:23.116  &  $-$05:56:10.97 \\
\hline
\end{tabular}
\end{table}

\subsection{Spitzer IRAC}
\label{sec:spitzer-irac}
The ONC has been extensively observed with {\it IRAC} on-board the {\it Spitzer} observatory in the course of programs 30641, 43 and 50070. We retrieved the corresponding IRAC BCD images and associated ancillary products from the public archive, and processed them following standard procedures with the recommended {\it MOPEX} software \citep{2005PASP..117.1113M}. The observations were all made using the High Dynamics Range mode, providing short (0.6~s) and long (12~s) exposure. We processed the two sets independently so as to cover the largest dynamic range. The procedure within {\it MOPEX} includes overlap correction, resampling, interpolation (to have an output pixel scale of 0\farcs6) and outlier rejection. The individual frames were then median combined using {\it Swarp} \citep{SWARP} using the rms maps produced by MOPEX as weight maps. Aperture photometry of all the sources brighter than the 3-$\sigma$ noise of the local background was extracted using \textit{SExtractor}. We verify that the corresponding photometry is in good agreement with IRAC photometry from  \citet{2009A&A...504..461F} within the uncertainties.

\subsection{Spitzer MIPS1}
\label{sec:spitzer-mips}
The ONC was observed with the {\it Spitzer Space Telescope} and its MIPS instruments in the course of programs 202, 30641, 30765, 3315, 40503, 47, 50070, and 58. We retrieved from the public archive all the corresponding MIPS1 (24~$\mu$m) BCD images, and processed them with the recommended MOPEX software. The procedure includes self-calibration (flat-fielding), overlap correction, outlier rejection, and weighted coaddition into the final mosaic. Aperture photometry of all the sources brighter than the 3-$\sigma$ noise of the local background was extracted using \textit{SExtractor}. We also verify that the corresponding photometry is in very good agreement with MIPS photometry from  \citet{2009A&A...504..461F} within the uncertainties.

\begin{figure*}[!tbp]
  \begin{center}
    \includegraphics[width=6in,angle=0]{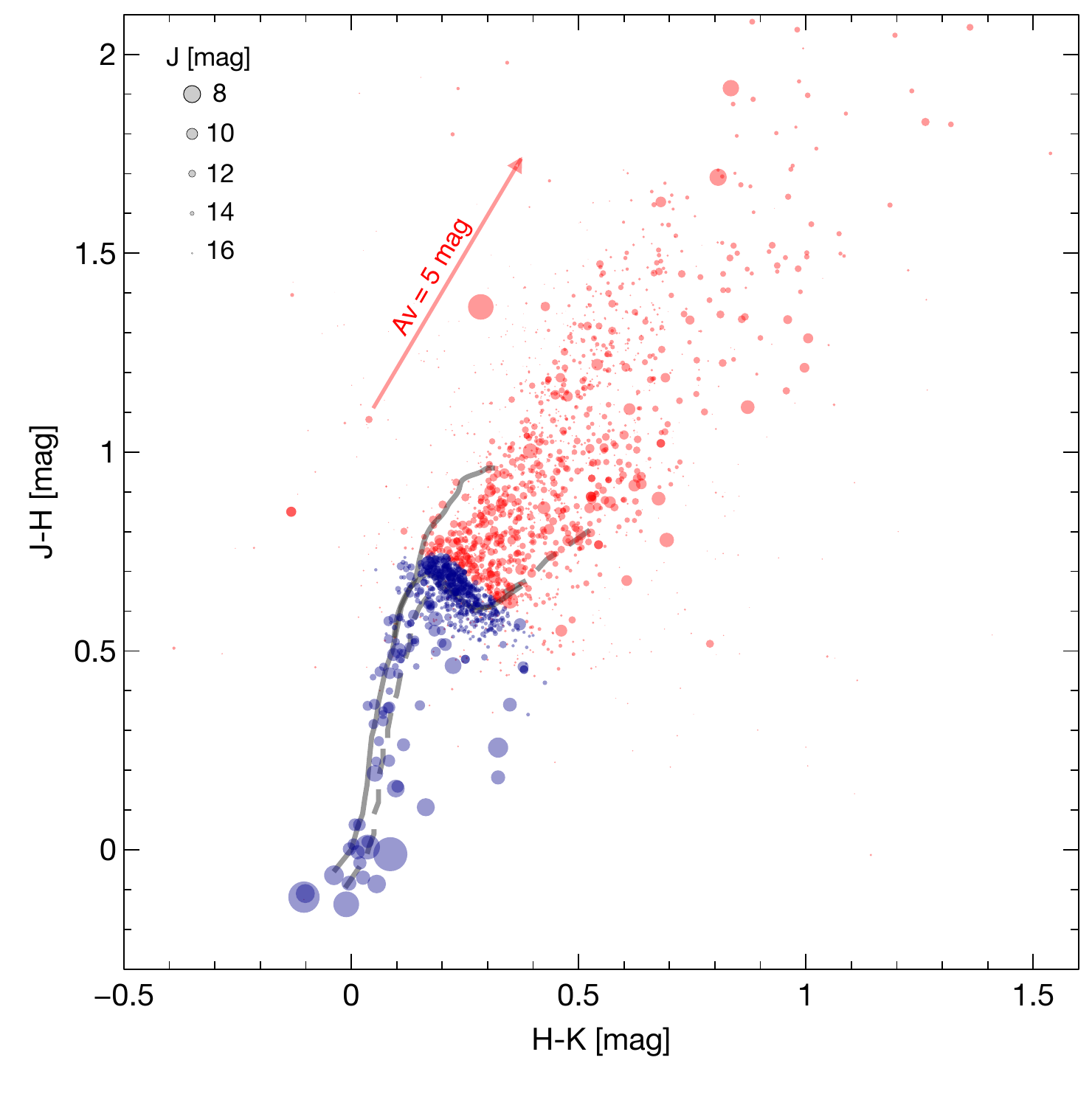}
    \caption{J$-$H versus H$-$K Color-Color versus J brightness diagram. The solid grey lines represent the main sequence and giant sequences from \cite{Bessel1988}. The dashed grey line represents the main sequence from \cite{2007AJ....134.2340K}. The sizes of the symbols are proportional to J-band brightness. Sources taken as foreground candidates are marked in blue, while rejected sources, mostly extincted sources, are marked in red. }
    \label{fig:jhkcc}
  \end{center}
\end{figure*}

\section{Results}
\label{sec:results}

We are interested in the foreground population to the Orion A cloud, in particular the foreground populations towards the ONC. To separate it from the background we will use the optical properties of dust grains in the Orion A cloud to block the optical light to the cloud background. This is a very effective way of isolating the stellar population between Earth and the Orion A cloud, in particular if we use blue optical bands where dust extinction is most effective.  To select the final sample of foreground stars we take two filtering steps informed by Color-Magnitude and Color-Color diagrams in the optical and infrared. In particular,  we start by 1) using blue optical magnitudes and colors to define a reliable subsample of sources in front of the cloud, then 2) using a near-infrared color-color diagram to reject sources affected by extinction (these are sources that are either young stars inside the cloud or background sources that are bright enough to be detected in the optical survey).

\begin{figure*}[!tbp]
  \begin{center}
    \includegraphics[width=\hsize,angle=0]{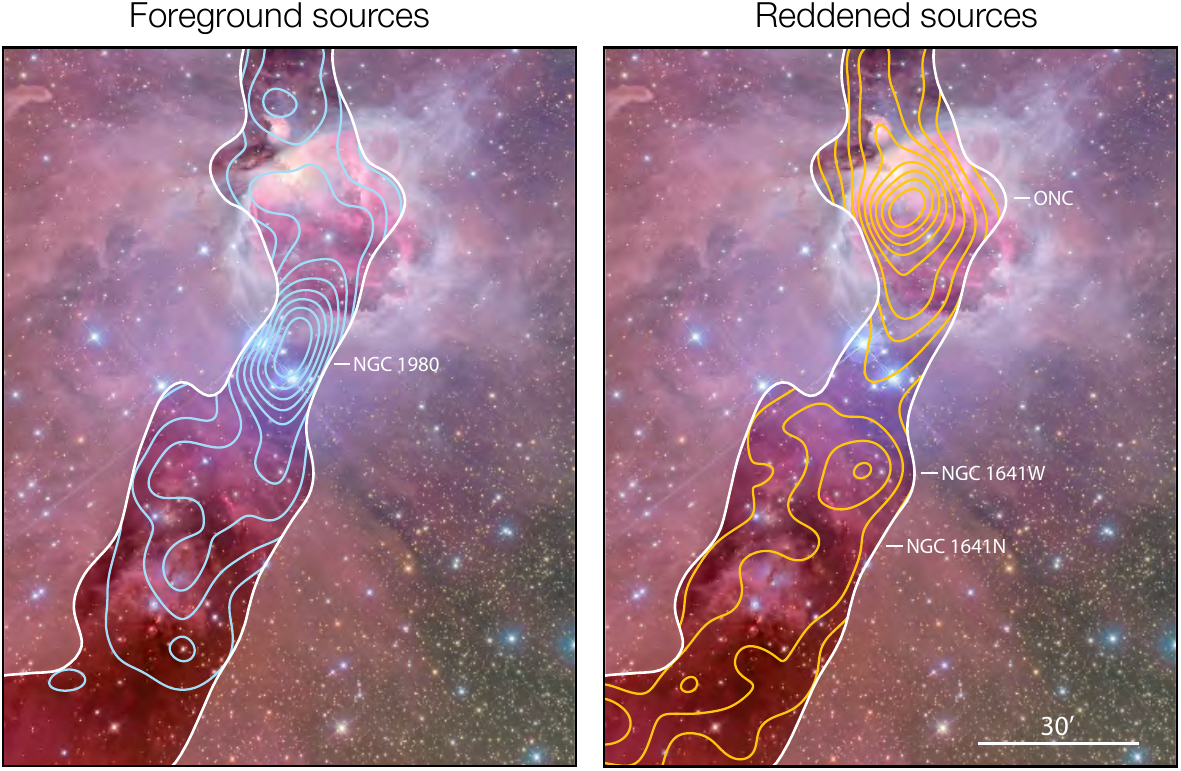}
    \caption{Left panel: spatial density of the foreground sources (blue sample). The unshaded area in the Figure represents the region of the cloud where A$_V \geq 5$ mag, on which the selection was performed. The blue contours (with increments of 10\% from the maximum) represent the surface density of foreground sources (constructed with a gaussian kernel with a width of 20\arcmin). Right panel: same as in the left panel but for the reddened sources. The  distribution of foreground sources shows a well defined peak coinciding with the poorly studied NGC 1980. The reddened sources, on the other hand, peak around 1) the Trapezium cluster and are mostly confined to the nebula and 2) the L1641N star forming region, with a peak towards a hitherto unrecognized group of YSOs (see text). The reddened and foreground populations are spatially uncorrelated but there is significant overlap between the two, in particular with the sources inside the Orion Nebula. 
}
    \label{fig:density}
\end{center}
\end{figure*}

\begin{figure}[!tbp]
  \begin{center}
    \includegraphics[width=\hsize,angle=0]{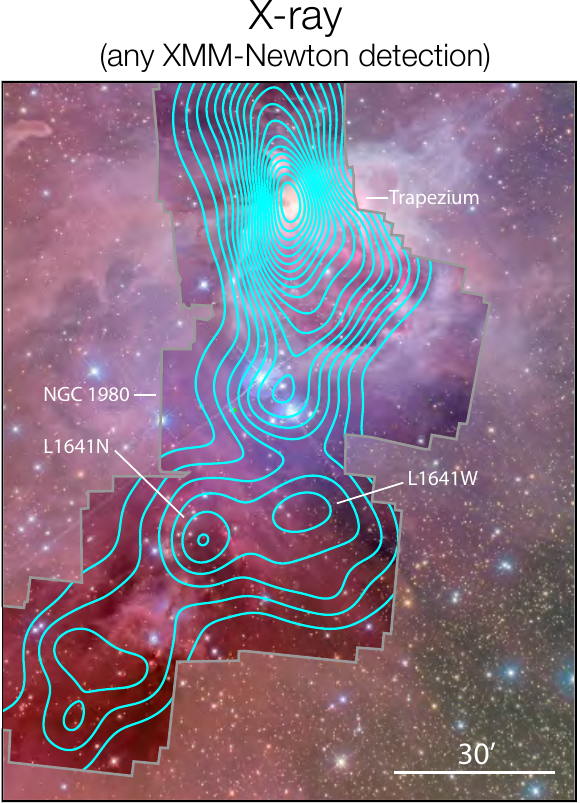}
    \caption{Spatial density of all the X-ray sources in the Third XMM-Newton serendipitous source catalog for the same region as in the previous Figure. The gaussian kernel used and contour separation are the same. NGC 1980 is detected as a distinct enhancement in the surface density of X-ray sources, together with the Trapezium cluster, the L1641N population, and the hitherto unrecognized group that we name L1641W.}
    \label{fig:density-xrays}
\end{center}
\end{figure}

The first filtering step is displayed in Figure~\ref{fig:grcmd} where we present a $g$ vs $g-r$ color-magnitude (CMD) diagram combining the SDSS, CFHT/Megacam and CAHA123 photometry. We chose $g-r$ over the more extinction sensitive $u-g$ as the $u-$band observations are significantly less deep, and extremely sensitive to excesses related to accretion and activity.

The first diagram (on the left) represents the $g$ vs $g-r$ color-magnitude diagram for all the sources in our combined database. The three CMD-diagrams to the right are a subset of the first containing only sources projected against increasing contours of dust extinction of the Orion A cloud (about 4, 5, and 6 magnitudes of visual extinction). These column density thresholds were estimated from the $^{13}$CO map of \citet{Bally87}, cross-calibrated with the extinction map of \cite{Lombardi2011}. While using directly the extinction map of \cite{Lombardi2011} gives similar results, we preferred to avoid dealing with any possible systematics affecting this map caused by a potential substantial population of foreground sources.  As we impose the condition of keeping only sources that are seen against increasing levels of dust extinction two things occur: 1) the number of stars decreases, naturally, because the solid angle on the sky decreases, and 2) a well defined sequence appears. This sequence is not what is expected from the general Galactic population between Earth and the Orion A cloud at 400 pc, as confirmed with the Besa\c{c}on stellar population model \citep{Robin03}. From this step we retain the subsample of sources that is seen in projection against column densities of greater than $\sim5$ visual magnitudes of extinction (third panel in Figure~\ref{fig:grcmd}), or a total of 2169 sources from more than 1.25$\times 10^5$ sources in the combined SDSS--MEGACAM--CAHA123 catalog. Most of the discarded sources have colors consistent with unreddened and slightly reddened unrelated field stars towards the background of the Orion A cloud. Among the sources that pass the first filter there could be some with g-band excess emission, but these should have a negligible effect in the selection process, in particular because the next filtering step is done at the near-infrared.

The second filtering step consists of discarding extincted sources. We want to remove from the  sample any source that might be associated with the cloud (young stellar objects still embedded in the cloud, for example), as well as background sources that are bright enough to be detected at $\sim$0.487$\mu$m (g-band) through A$_V\sim5$ mag of cloud material. We perform this filtering using a J$-$H vs. H$-$K Color-Color diagram  where extincted sources are easily identified along the reddening band, away from their unreddened main sequence (and giant) colors \citep[e.g.][]{1992ApJ...393..278L,1998ApJ...506..292A,2001A&A...377.1023L}. We present in Figure~\ref{fig:jhkcc} the J$-$H vs. H$-$K Color-Color diagram for the 2169 sources that passed the first filtering step. The size of the symbols in this Figure are proportional to J-band brightness. The selection criteria used to identify the likely foreground population was:

\begin{equation}
  \label{eq:1}
  H < \frac{0.96-(J-H)}{1.05} \,\, \mathrm{mag}
\end{equation}

\begin{equation}
  \label{eq:2}
  J < 15 \,\, \mathrm{mag}
\end{equation}

\begin{equation}
  \label{eq:3}
  J-H<0.74 \,\, \cup \,\, H-K>-0.2 \,\, \cup \,\, H-K<0.43  \,\, \mathrm{mag}
\end{equation}

\noindent 
Sources that are consistent with having no extinction within the photometric errors, are marked in blue, while rejected sources are marked in red. Condition (1), the main filter, is taken as the border between extincted and non extincted sources, and it was selected to be roughly parallel to the main-sequence early M-star branch (to about the color of a M4-M5 star). Condition (2) and (3) further reject sources that are faint or have dubious NIR colors (either bluer than physically possible, or redder than main-sequence stars, suggestive of a NIR excess). Condition (2) in particular makes the selection more robust against photometric errors (the typical photometric error imposed by condition (2) is J$_{err} \sim$ 0.03$\pm$0.01, H$_{err} \sim$ 0.03$\pm$0.01, and K$_{err} \sim0.03\pm0.01$ mag, which translates into a maximum A$_V$ error of $\sim$ 1 mag), and should reach a sensitivity limit capable of including M4 main-sequence stars at the distance of the cloud (J $\sim$ 15 mag). More than two thirds of the 2169 sources are rejected (red sources) and a total of 624 sources have colors consistent with foreground stars suffering no or negligible amounts of extinction.\footnote{A table of candidate foreground sources is provided in electronic format.}

\begin{figure}[!tbp]
  \begin{center}
    \includegraphics[width=\hsize,angle=0]{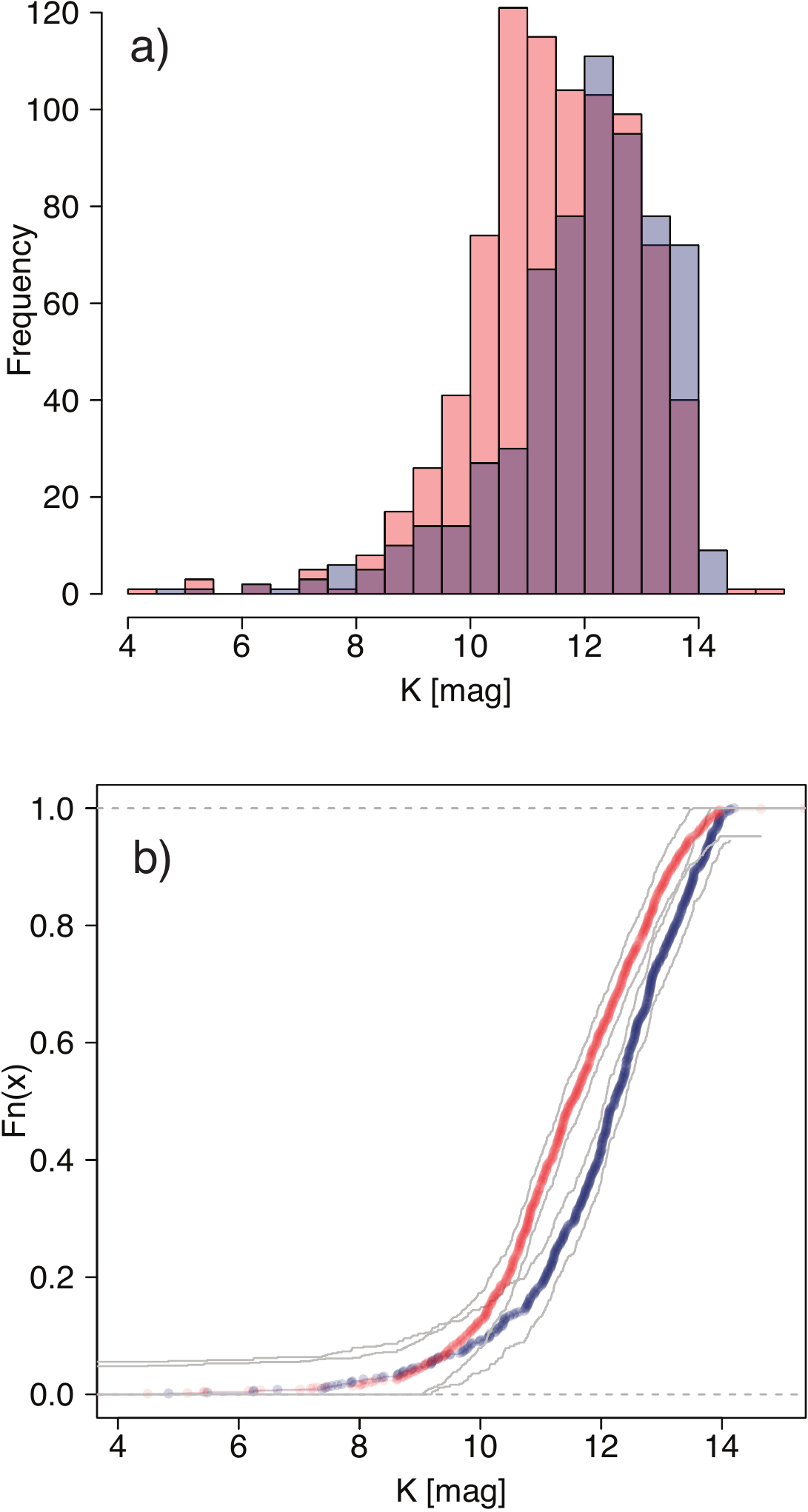}
    \caption{a) K-band luminosity functions  for the foreground (blue) and the extincted (red) sample (see Figure~\ref{fig:jhkcc}). Albeit being affected by extinction, the red sample is surprisingly  brighter than the  foreground sample (blue), suggesting that the two populations are intrinsically different. b) Empirical cumulative distribution functions for both samples, together with upper and lower simultaneous 95\% confidence curves, confirming that the two populations have statistically different luminosity functions.
 }
\label{fig:klf}
\end{center}
\end{figure}

\subsection{Spatial distribution of the foreground and reddened sources}
\label{sec:spatial-distribution}

The addition of the third dimension (J-band brightness) to Figure~\ref{fig:jhkcc} makes obvious the presence of a well populated main-sequence branch, from early types (B-stars) to late types (M-stars). The clear presence of so many early-type stars, as well as a well defined main sequence, suggests that the foreground population is not dominated by the random Galactic field from Earth to the ONC region, but could  instead be a well defined stellar population. To investigate this idea we present in the left panel of Figure~\ref{fig:density} the spatial density of foreground sources (blue sample). The unshaded area in the Figure represents the region of the cloud where A$_V \geq 5$ mag, on which the selection was performed. We constructed a surface density map of foreground sources using a gaussian kernel with a width of 20\arcmin\, represented in the Figure as blue contours (with increments of 10\% from the maximum). It is clear from Figure~\ref{fig:density} that the distribution of foreground sources is not uniform, as expected for the Galactic field between Earth and the ONC region in the Orion A cloud, but is instead strongly peaked and fairly symmetric. The peak of the distribution coincides spatially with the poorly studied iota Ori cluster, or NGC 1980, suggesting that the foreground population is dominated by NGC 1980 cluster members. The elongated shape of the peak is not meaningful as it is caused by the relative narrow A$_V \geq 5$ mag region on which the density was calculated. 

The right panel of Figure~\ref{fig:density} shows the spatial distribution of reddened sources. This distribution is dominated by two peaks, a relatively well defined one in the Trapezium cluster region and another, more diffuse, coinciding with the overall gas distribution around the L1641N star forming region. This is not surprising as the reddened sources are expected to be dominated by the embedded young stellar objects in the Orion A cloud, and these are know to cluster around these two regions.  More striking, instead, is that the foreground sources and the reddened sources appear spatially anti-correlated: the maximum of the distribution of the foreground sources coincides with the minimum of the distribution of reddened sources. This is evidence that the foreground population is not the emerging young stellar population from the ONC but is instead an entirely different population. Because it contains a fully sampled and unreddened main-sequence from B- to M-stars (see Figure~\ref{fig:jhkcc}), this population is most likely the stellar population of the NGC 1980 cluster, seen in projection against the Orion A cloud. But because this population overlaps significantly with the ONC (the distribution of foreground sources appears symmetric to about 7 pc from its center), it implies that the ONC is not comprised by a single cluster with the Trapezium as its core, but has instead three stellar populations (a) the youngest population, including the Trapezium and ongoing star formation associated with the dense gas in the nebula, b) part of the NGC 1980 cluster in its foreground, and c) the unrelated, Galactic, foreground and background population. 

In Figure~\ref{fig:density-xrays} we present, for the same region shown in Figure~\ref{fig:density}, the distribution of all the X-ray sources in the Third XMM-Newton serendipitous source catalog. We want to investigate if the NGC 1980 X-ray source counts appear as a distinct peak in the surface density map (constructed  using a gaussian kernel with a width of 20\arcmin\, as in the previous Figure). As can be seen in this Figure, both the foreground and the extincted populations are detected, and a distinct enhancement in the surface density of X-ray sources is seen towards the center of NGC 1980, given strength to the idea that the foreground population is not the emerging young stellar population from the ONC but is instead an entirely different population. The highest peak in this surface density map is centered on the Trapezium cluster. This was not the case for the extincted population as seen in Figure~\ref{fig:density}, which peaks slightly to the South of the Trapezium, although this mismatch could be due simply to the fact that our MEGACAM g-band observations are affected by the bright nebula and dust extinction in the Trapezium region, hence substantially less sensitive to the embedded population of the Trapezium cluster.

\subsubsection{A hitherto unrecognized group of YSOs?}
\label{sec:hith-unrec-group}

The presence of an enhancement of the reddened sources in Figure~\ref{fig:density} towards RA: $5^h35^m$, Dec: $-6^\circ18^m$, immediately South of NGC 1980 and towards the West of what is normally taken as the L1641N cluster \citep[e.g.][]{Allen2008}, is tantalizing. Could this relatively small enhancement be another hitherto unrecognized group of YSOs?  The enhancement is clearly detected in X-rays (Figure~\ref{fig:density-xrays}), and is tentatively detected in the optical (Figure~\ref{fig:density}), providing support in favor of this possibility. We name this potentially new group of about 50 stars (counted on the reddened sample) as L1641W. The group is not associated with any obvious nebula nor does it include any obvious bright star. Because it appears less extincted than the L1641N population, and it is not obviously detected in the Spitzer survey \citep[e.g.][]{Allen2008}, suggests that it is probably more evolved than the L1641N population.  We speculate that this new group is either a foreground young group ramming into the Orion A cloud, or a slightly older sibling of NGC 1641N, leaving the cloud.

\begin{table}
\caption{Position of clusters in Figure~\ref{fig:density-xrays}, including the newly identified L1641W  \label{clusters}}
\begin{tabular}{lcc}\hline\hline
 Field          &  RA (J2000)        & Dec (J2000)             \\
\hline
L1641W      & 05:34:51.0  &  $-$06:17:40 \\
NGC 1980  & 05:35:11.0  &  $-$05:58:00 \\
Trapezium  & 05:35:16.5  &  $-$05:23:14 \\
L1641N       & 05:35:55.7  &  $-$06:23:55 \\
\hline
\end{tabular}
\end{table}

\subsection{Luminosity function of foreground and reddened sources}
\label{sec:lumin-funct-blue}

In Figure~\ref{fig:klf} a) we present the K-band luminosity functions for both the foreground (blue) and reddened (red) samples (see Figure~\ref{fig:jhkcc}). To enable a direct comparison, the reddened sample was also constrained with condition (2), namely, J $<$ 15 mag.  Surprisingly, the extincted sample (red) is brighter than the foreground sample (blue), even if no derredening procedure was applied to the red sample.  We confirm that the differences between the two luminosity functions are significant, to a 95\% confidence level, by analyzing their empirical cumulative distribution functions (see Figure~\ref{fig:klf} b)). Note that had we de-reddened the extincted sample, the difference between the luminosity functions would have been even higher. This suggests, like in the previous section, that the foreground and reddened population are intrinsically different. A likely explanation for the difference in the luminosity functions is that the reddened sample is dominated by very young stellar objects still embedded the cloud, which are intrinsically brighter than normal stars because of both stellar evolution and the presence of K-band excess emission.

\begin{figure}[!tbp]
  \begin{center}
    \includegraphics[width=\hsize,angle=0]{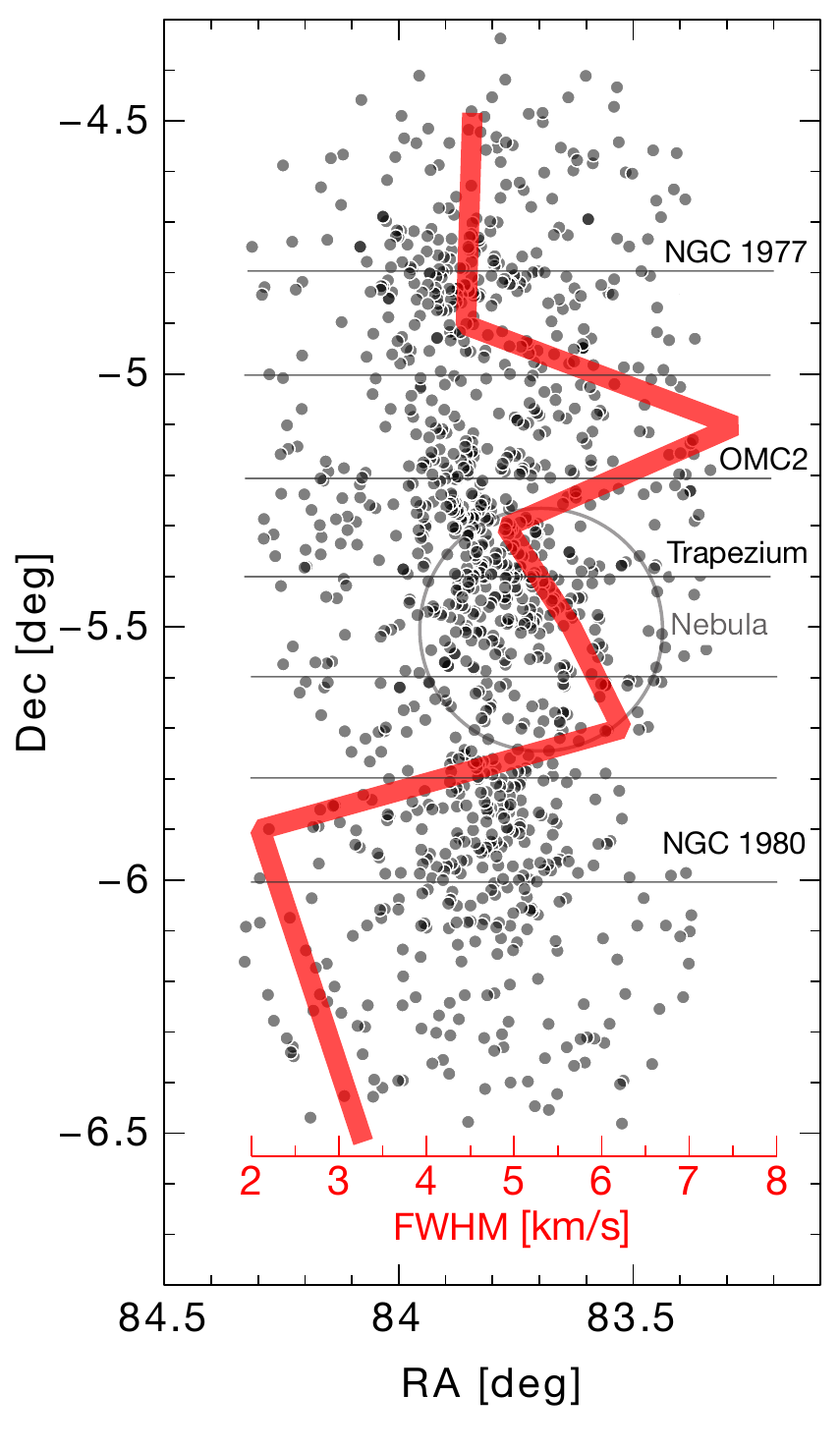}
    \caption{The North-South velocity dispersion profile of the ONC region. The filled circles represent the sources in \citet{2009ApJ...697.1103T} with reliable radial velocities. The NGC 1977, the Trapezium cluster, and NGC 1980 are indicated, as well as the extent of the Orion Nebula (light open circle).  The thick red line represents the North-South velocity dispersion profile measured in bins of Declination (indicated by the thin horizontal lines). There is  an increase from the Trapezium to the edge of the nebula to the South, followed by a clear minimum around NGC 1980, strongly suggesting that NGC 1980 is a well defined and different population from the stellar population inside the nebula. Data from Table 13 of \citet{2009ApJ...697.1103T}.}
    \label{fig:vdispprofile}
  \end{center}
\end{figure}

\subsection{Velocity dispersion profile}
\label{sec:veloc-disp}

\citet{2009ApJ...697.1103T} presented a large kinematic study of the ONC, covering about 2$^\circ$ of Declination centered on the Orion A cloud, from NGC 1977 down to L1641N. This survey builds on previous work by \citet{2008ApJ...676.1109F} and constitutes the largest and highest precision kinematic survey of this region to date, offering a unique possibility to characterize kinematically the ONC foreground population identified in this paper. In Figure~\ref{fig:vdispprofile} we present the North-South velocity dispersion profile of the ONC region, taken from Table 13 of \citet{2009ApJ...697.1103T}. The filled circles represent the sources in this paper with reliable radial velocities. The NGC 1977, the OMC2/3, the Trapezium cluster, and NGC 1980 are indicated, as well as the extent of the Orion Nebula (light open circle).  The thick red line represents the North-South velocity dispersion profile measured in bins of Declination (indicated by the thin horizontal lines).  

It is striking that the velocity dispersion profile has a minimum at the location of NGC 1980. This is perhaps the strongest indication we have that the stellar population of NGC 1980 is a distinct population from the reddened population inside Orion A. The measurement of velocity dispersion in the bin that mostly includes NGC 1980 ($\sigma=$2.1 km/s) was not optimized to isolate the most probable members of this cluster, and should then be seen as an upper limit to the true velocity dispersion in this cluster. Still, this value is close to the velocity dispersion of the Trapezium cluster as measured from the proper motion of stars within one half degree of the center of the Trapezium, namely, 1.34$\pm$0.18 km/s for a sample of brighter stars \citep{1988AJ.....95.1744V} and 1.99$\pm$0.08 km/s for a larger sample including relatively fainter stars \citep{1988AJ.....95.1755J}. Both velocity dispersions were corrected to the more recently estimate of the distance to the Orion A cloud (400 pc). 

Given the striking differences in the velocity dispersion profile we then calculated the mean radial velocity per bin from the subsample of single sources (not directly available in the Tobin et al. 2009 paper) and found that although showing variations from bin to bin, these variations are of the order of the measured dispersions. In particular, the mean velocity for the bin including the Trapezium ($-5.3^\circ < \delta < -5.4^\circ$) and NGC 1980  ($-5.8^\circ < \delta < -6.0^\circ$) is $25.7\pm3.0$ km/s and $24.3\pm2.7$ km/s respectively. Within the errors, estimated as the median absolute deviation in each bin, the NGC 1980 cluster has virtually the same radial velocity as the ONC. We note, however, that we are taking the bins as simple slices at constant declination, without trying to optimize their boundaries to better separate the different populations. 

Because of the importance of measuring the velocity differences between the Trapezium and NGC 1980, especially for a discussion on the origin of NGC 1980, we made an alternative source selection and created two new subsamples, that are in principle more pure, but have about three times less sources. For the NGC 1980 subsample we matched the Tobin et al. 2009 catalog with the foreground population identified in this paper. For the Trapezium we matched the Tobin et al. 2009 catalog with the COUP sample \citep{2002ApJ...574..258F}, that is dominated by Trapezium sources, and removed sources that matched the foreground population. Because this Trapezium subsample should be of ``high confidence'', we used the radial velocity limits found in this subsample (6.2 km/s and 36.6 km/s) to exclude 5 extreme outliers in the NGC 1980 sample (with velocities of $\sim -40$ and $\sim 90$ km/s). In these subsamples, the mean velocity for the Trapezium and NGC 1980 clusters is $25.4\pm3.0$ km/s and $24.4\pm1.5$ km/s respectively, or essentially the same values as derived above, with the important difference that the dispersion of velocities in NGC 1980 is now reduced by about a factor of two, once again suggesting that this cluster is a distinct population from the reddened population inside Orion, as argued above. Still even with the decreased velocity dispersion the measured velocity difference of 1 km/s is not statistically significant.

\subsection{On the age and population size of NGC 1980}

\begin{figure}[!tbp]
  \begin{center}
    \includegraphics[width=\hsize,angle=0]{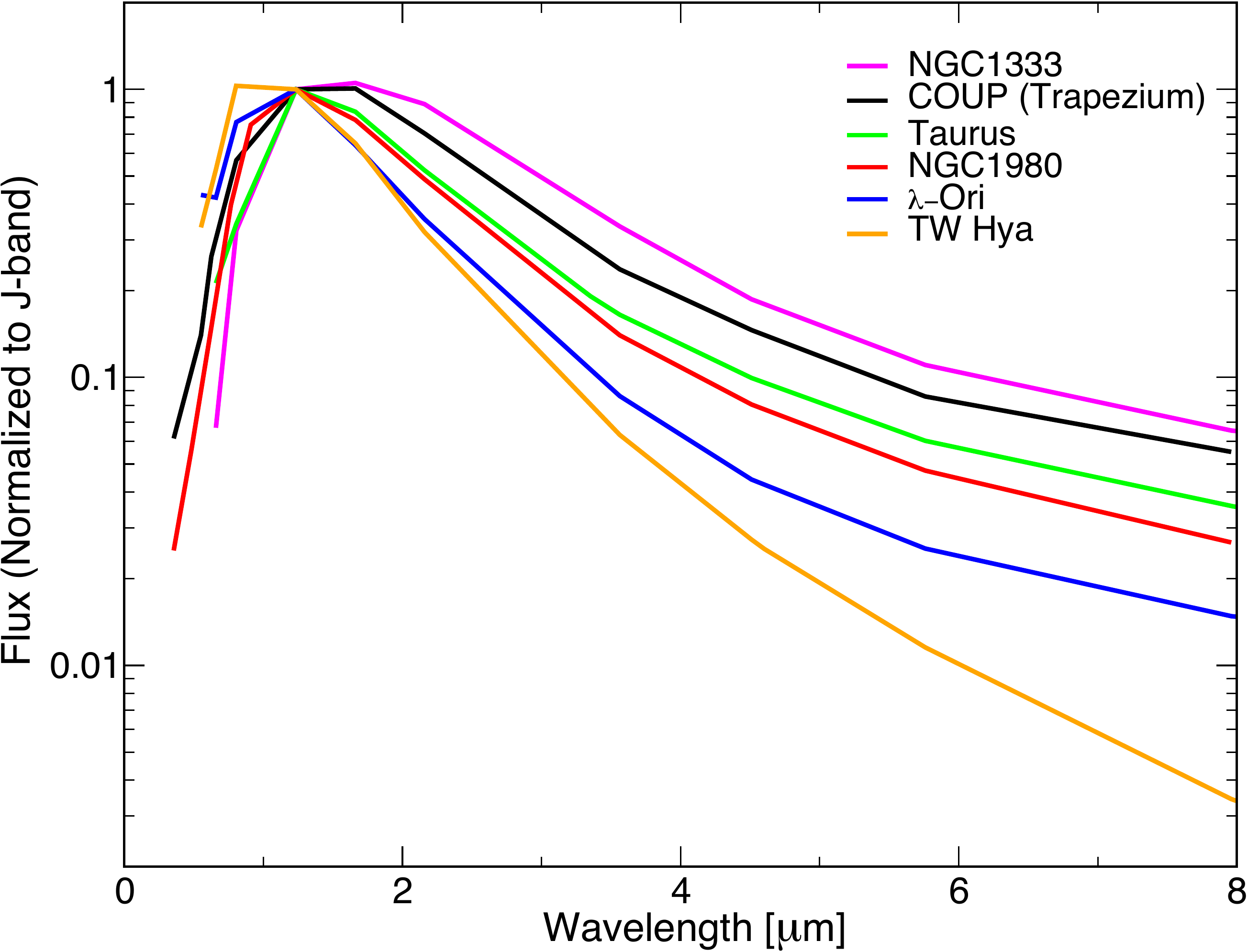}
    \caption{Median SED of class~II sources in clusters of various evolutionary stages: NGC1333 (magenta, 1~Myr), Trapezium COUP sources (black, 1~Myr), Taurus (green, 1--3~Myr), NGC1980 members (red), $\lambda$ Ori (blue 5--7~Myr) and $\eta-$Cha (orange, 5--10~Myr).}
    \label{fig:medsed}
  \end{center}
\end{figure}

In order to estimate an age to the NGC 1980 cluster we compare the evolutionary status of class~II sources in various clusters through the analysis of the median spectral energy distribution (SED) of late-type (spectral type later than K0) members. We follow the \citet{2005ApJ...629..881H} definition of Class II, namely objects with $0.2<[3.6]-[4.5]<0.7$~mag and $0.6<[5.8]-[8.0]<1.1$~mag. To compute the median SED for the different clusters we retrieved the optical, near-infrared (2MASS) and mid-infrared ({\it Spitzer} and {\it WISE}) photometry for samples of confirmed members of Taurus \citep[1--3~Myr,][]{2010ApJS..186..111L}, IC~348 \citep[1--3~Myr,][]{2006AJ....131.1574L}, NGC1333 \citep[1~Myr,][]{2010AJ....140..266W,2008ApJ...674..336G}, $\lambda-$Ori \citep[5--7~Myr,][]{2001AJ....121.2124D,2004ApJ...610.1064B}, and $\eta-$Cha \citep[5--10~Myr,][]{2005ApJ...634L.113M}. To compute the median SED for the Trapezium cluster we defined first a ``high confidence'' Trapezium member catalog, as we did in the previous Section, by cross-matching the X-ray COUP sample from \citep{2008ApJ...677..401P} with the foreground (NGC 1980) sample, and excluding all matches as unrelated foregrounds.  The individual SEDs within each cluster were normalized to the $J$-band flux, and the median cluster SED of each cluster was computed. Figure~\ref{fig:medsed} shows the result. One can see from this Figure~\ref{fig:medsed} that the optical part of the SED varies from cluster to cluster, mostly due to dust extinction. More striking, the mid-infrared ($>3$~$\mu$m) excesses, related to the presence of a disk, decrease systematically with age.

The median SED of NGC 1980 seems to fit between the median SED of Taurus (1--3~Myr) and $\lambda-$Ori (5--7~Myr), suggesting an age in between that of these regions. But another constraint is given by the massive stars in the center of the cluster. Of the five brightest stars at the peak of the spatial distribution in Figure~\ref{fig:density}, only the brightest, iota Ori (O9 III, V$=$2.77 mag), seems to have evolved from the main sequence. This implies an age of about 4-5 Myr for this star, assuming it started its life as a 25 M$_\odot$ star \citep[e.g.][]{Massey:2003fk}. This age fits well within the inferred age from the median SED and is also in agreement with the estimate of \citet{1978ApJS...36..497W} for the age of Ori OB 1c subgroup (of about 4 Myr).   

To estimate the size of the cluster population we concentrate on the distribution of foreground sources from the center of the cluster to the South in order to avoid incompleteness issues caused by the bright Orion Nebula. We counted the number of sources falling on a 20$^\circ$ ``pie slice'' inside the A$_V \geq 5$ mag region, centered on the cluster and with a radius of 7 pc. This radius corresponds approximately to the extent of the 10\% contour in Figure~\ref{fig:density}, chosen to account for contamination from the Galactic field between Earth and Orion (estimated to be 6--9\% of the foreground population in Section~\ref{sec:unrel-galact-field}). Note that this radius is not the half-mass radius but simply the radius to which we can trace the enhancement of sources over the unrelated foreground field.
 We repeated this measurement several times to account for uncertainties in the location of the cluster center and obtained an average number of 110 sources in the  20$^\circ$ ``pie slice''. Assuming spherical symmetry for the distribution of sources in NGC 1980, we expect then a total of about 2000 sources in NGC 1980, or a total cluster mass of about 1000 M$_\odot$ (assuming an average mass per star of 0.5 M$_\odot$). 

Assuming NGC 1980 has a normal Initial Mass Function (IMF), we can make a consistency test on the likelihood of the number of sources in this cluster being of the order of 2000. For this we constructed 200000 synthetic clusters of 1000 M$_\odot$ each by randomly sampling the Kroupa and the Chabrier IMFs \citep{Kroupa2001,2003PASP..115..763C} and tracked the mass of the most massive star in each synthetic cluster. The mean mass of the most massive star was 54$\pm$26 M$_\odot$ (Kroupa) and 22$\pm$11 M$_\odot$ (Chabrier). Assuming there were no supernovae in NGC 1980 yet, and that iota Ori is the most massive star in the cluster then, to first approximation, a population of about 2000 sources seems a plausible estimate of the size of NGC 1980 population.

At the moment we cannot derive a reliable cluster radial profile, nor a half-mass radius, or even be certain about the position of the center of the cluster, as our optical observations are incomplete in the vicinity of the early type stars of NGC 1980, and we only have a ``pie slice'' view on the radial extent of the cluster. This should be improved in follow-up work, in particular in combination of dedicated NIR observations which are less sensitive to the large brightness contrasts between early and late type stars in this cluster.

\begin{figure*}[!tbp]
  \begin{center}
    \includegraphics[width=\hsize,angle=0]{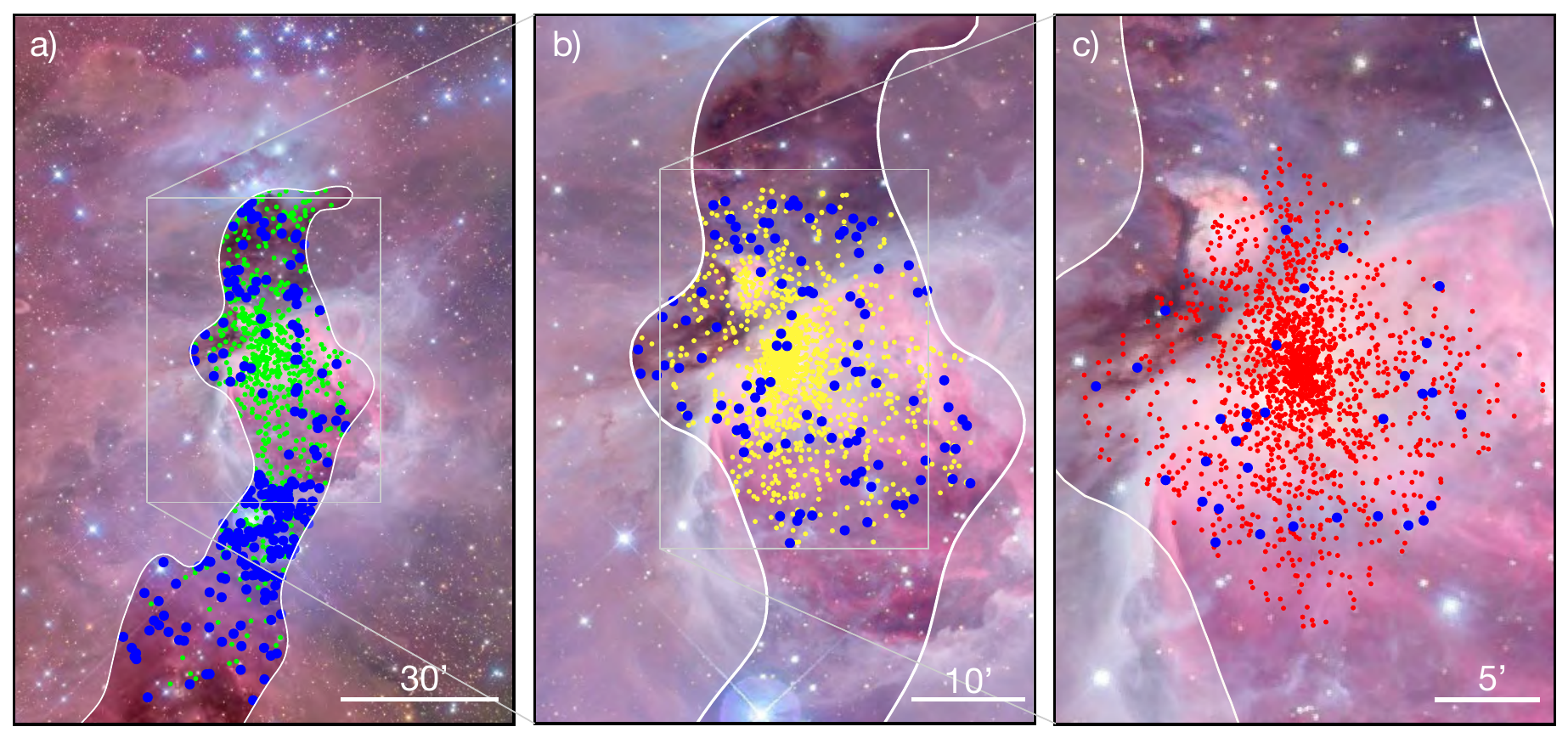}
    \caption{Matches between the foreground contamination and well known catalogs of the ONC. a) In green, the optical spectroscopic survey of \cite{2009ApJ...697.1103T}, with the matches with the foreground population marked in blue. b) In yellow, the well studied sample of \cite{Hillenbrand1997}, with the matches with the foreground population marked in blue. c) In red, the sample of X-ray sources from the COUP project \citep{2002ApJ...574..258F}, with the matches with the foreground population marked in blue. As in Figure~\ref{fig:density} the unshaded area represents the region of the cloud where A$_V \geq 5$ mag, on which the selection of foreground sources was performed. }
    \label{fig:contamination}
  \end{center}
\end{figure*}

\subsubsection{On the origin of NGC 1980 and impact on the Orion A cloud}
\label{sec:possible-impact-ngc}

We found in Section~\ref{sec:veloc-disp} that the radial velocity of NGC 1980 is indistinguishable, or has a difference of the order of a few km/s at best from the radial velocity of the embedded Trapezium population. This surprising result implies that the radial velocity of NGC 1980 is essentially the same as the velocity of the gas in the Orion A cloud, since the ONC population has the same radial velocity as the cloud \citep{2009ApJ...697.1103T}. This strongly suggests that NGC 1980 is somehow connected to the Orion A cloud, or better, that the cloud that formed NGC 1980 was physically related to the current Orion A cloud. One would not expect the distance to NGC 1980 to be substantially different than the current distance estimate to the ONC, and a fitting of the ZAMS on the optical data presented in this paper is indeed consistent with a distance of 400 pc. 

Despite its relatively older age, lack of obvious H$_{\rm{II}}$ region, and lack of measurable dust extinction, NGC 1980 moves away from Earth at the same velocity as the large Orion A cloud on which it is seen in projection. Because of their likely proximity, one wonders on the effects of the ionizing stars from NGC 1980 on the Orion A cloud, or what the cloud was about 4-5 Myr ago, in particular on a possible acceleration and compression of the cloud by the UV radiation from these stars. How important was/is this process in this region? Could the formation of the ONC have been triggered by its older sibling, as suggested in \citet{Bally2008}?  At first look our results would argue that the impact would have been minimal, NGC 1980 has essentially the same radial velocity as the Orion A cloud, but the work of \citet{1954BAN....12..177O,1954BAN....12..187K,Oort:1955cn} suggests that final speeds between the ionizing star and the cloud would be of the order of a few km/s, which cannot be ruled out by the current accuracy of the data. While our results do not give final evidence in support of the tantalizing suggestion that the formation of the ONC could have been triggered by NGC 1980, they are not inconsistent with it either. A new dedicated radial velocity survey of the region, together with a sensitive proper motion survey, are needed to understand the interplay between these two massive clusters. This configuration (an embedded cluster in the vicinity of a $\sim 5$ Myr cluster) is unlikely to be unique in massive star forming clouds, but it will best addressed in the nearest example.

\subsection{Contamination of ONC catalogs}
\label{sec:cont-catalogs}

We have showed above that there is a rich and distinct foreground population of stars, likely associated with the young ($\sim 5$ Myr) poorly studied but massive NGC 1980 cluster, that is not directly associated with the ongoing star formation in the ONC. This finding raises concerns on the contamination of currently available observables for this important region, and future studies should take this foreground population into account. But how large is this contamination? There are  two well known ONC catalogs used in the literature, namely the \citet{Hillenbrand1997} and the catalogs of \citet{2009ApJS..183..261D,DaRio:2010cz,2012ApJ...748...14D} covering a roughly square area of about $0.5^\circ\times0.5^\circ$ ($\sim 3.5 \times 3.5$ pc) centered on the Trapezium cluster. The  \citet{2012ApJ...748...14D} supersedes all previous catalogs, but it is the most recent hence less used in the community. On the other hand, the \citet{Hillenbrand1997} catalog has been used extensively in the literature and has spawned a large number of the star formation studies on the star formation properties of the ONC region. We estimate here the likely foreground contamination fraction for the \citet{Hillenbrand1997} catalog as it is the most used one, but also because it is likely to be the least contaminated since the Da Rio catalogs cover a slightly larger area of the sky towards NGC 1980.

To estimate the probable contamination fraction of \citet{Hillenbrand1997} we matched the foreground population with this catalog for stars falling within the A$_V \geq 5$ mag region where the foreground was selected (see Figure~\ref{fig:density}) and where I-band $<$ 16 mag. The last constraint accounts for the fact that the \citet{Hillenbrand1997} sample is not uniformly deep (it reaches about 2 magnitudes deeper around the Trapezium cluster), and that the selection of foreground stars, made at g-band, seems complete to about I-band $\sim$ 16 mag (after transformation of the SDSS photometry into Johnson's \citep{2007AJ....134..973I}). We find that 11\% of the sources in the \citet{Hillenbrand1997} catalog have a match in the foreground sample (8\% if we remove the constraint on the I-band brightness). If one sees the Trapezium cluster as a component of the ONC, and not as the only component, and remove it from consideration, then the fraction of foreground contaminants in the ONC rises to 32\%. For this estimate the area on the sky covered by the Trapezium cluster is taken as an ellipse with a$=7.5^\prime$ and b$=3.8^\prime$, with a position angle of $-10^\circ$, similar to the definition in \citet{Hillenbrand1998}. One can also make an estimate of the possible contamination to the entire \citet{Hillenbrand1997} catalog by applying equation~\eqref{eq:1} to the entire \citet{Hillenbrand1997} catalog, which gives then a contamination fraction of 20\%, or 63\% when the Trapezium is removed from consideration. Note that all these estimations assume that the fraction of ONC stars without measurable extinction is negligible, which is likely given the distribution of foreground stars in Figure~\ref{fig:density}, but will need to be investigated further in future work.

Even without removing the Trapezium cluster from consideration, contamination fractions of the order 10--20\% are significant and will necessarily lead to systematic errors in the basic derived physical quantities for this star formation benchmark. Still, these are necessarily lower limits to the true contamination fraction of the ONC sample for at least two reasons: 1) our g-band MEGACAM survey is not as sensitive in regions of high nebula brightness, especially around the Trapezium, and 2) we are not sensitive, by design, to background sources. While it is normally argued that the high background extinction behind the Trapezium blocks most background stars, this is only valid for the inner regions of the Trapezium cluster  ($\sim25^{\prime \,2}$), but not valid for the entire $\sim700^{\prime \,2}$ Orion Nebula, \citep[e.g.][]{1999ApJ...510L..49J,2012MNRAS.422..521B}. So background contamination is variable across the ONC and expected for any optical or infrared  survey of this region. In regards to the contamination of the ONC region by NGC 1980, it is a function of the position in the Nebula, having a minimum at the center of the Trapezium where the Trapezium cluster stellar density is highest, increasing gradually towards the South as one approaches the core of NGC 1980 (see Figure~\ref{fig:density}). 

\subsubsection{Unrelated Galactic field foreground population}
\label{sec:unrel-galact-field}

Most of the identified foreground population (624 sources) is likely to belong to NGC 1980, as seen from the symmetric and peaked spatial distribution in Figure~\ref{fig:density}, but some fraction of these must be made of the Galactic field population between Earth and the Orion A cloud. A first and simple estimate of the size of this population can be made by correlating the foreground population with the sample of \citet{2009ApJ...697.1103T} for which good radial velocity measurements exist (see Figure~\ref{fig:contamination}). The distribution of radial velocities of the 188 sources in common reveals a gaussian like distribution centered at $\sim 26.1$ km/s, with a gap roughly between $-20$-$0$ km/s and $40$-$60$ km/s without any source, and 3 sources below $-20$ km/s and 9 above 60 km/s (i.e., 12 potential outliers). Not surprisingly, if we sigma-clip the entire distribution at 3 $\sigma$ we find 11 outliers. If we remove from the distribution the 12 potential outliers described above and then sigma-clip the rest of the distribution we find 5 more outliers, but this time at the wings of the distribution. This suggests that about 6--9\% of the foreground sources identified in this work are likely field sources unrelated to NGC 1980. This estimate of the Galactic field foreground contamination for Orion A is in rough agreement with what would be expected from the Besa\c{c}on stellar population model \citep{Robin03} for the depth of our MEGACAM survey.

\section{Conclusions}

We have made a link between the foreground population towards the well known star formation benchmark ONC region and the stellar population of the poorly studied NGC 1980 cluster (or iota Ori cluster). Not only did we detect the presence of a well populated main-sequence (from B-stars to M-stars), the foreground sources have: 1) a well defined spatial distribution peaking near iota Ori, 2) a fainter luminosity function when compared to the extincted young population embedded inside the cloud, and 3) a low velocity dispersion, typical of that of other young clusters. 

Unlike the ONC, NGC 1980 is a relatively older cluster ($4-5$ Myr), lacks an obvious H$_{\rm{II}}$ region, and is comparatively free of dust extinction. Surprisingly, the radial velocity of NGC 1980 is currently indistinguishable from the radial velocity of the ONC embedded population or the radial velocity of Orion A cloud, suggesting that both clusters are genetically related, and at about the same distance from Earth.

A general concern that this study raises is the risk of population mixing in star formation studies. It is unlikely that the ONC is atypical in this respect and a dedicated multi-wavelength study to disentangle the different populations, together with a sensitive proper motion survey of the region, is urgently needed. The ONC is still the closer massive star formation region to Earth, and albeit more complicated than first assumed, it is still the one offering the best detailed view on the formation of massive stars and clusters.

The main conclusions of this work are:

\begin{itemize}

\item We make use of the optical effects of dust extinction to block the background stellar population to the Orion A cloud, and find that there is a rich foreground stellar population in front of the  cloud, in particular the ONC. This population contains a well populated main-sequence, from B-stars to M-stars.
 
\item The spatial distribution of the foreground population is not random but clusters strongly  around NGC 1980 (iota Ori), has a fainter luminosity function, and different velocity dispersion from the reddened population inside the Orion A cloud. This foreground population is, in all likelihood, the extended stellar content of the poorly studied NGC 1980 cluster.

\item We estimate the number of members of NGC 1980 to be of the order of 2000, which makes it one of the most massive clusters in the entire Orion complex, and estimate its age to be $\sim 4-5$ Myr by making a comparative study of median spectral energy distributions among known young populations and constraints from the age of post main sequence star iota Ori.  

\item This newly found population overlaps significantly with what is currently assumed to be the ONC and the L1641N populations, and can make up for more than 10-20\% of what is currently taken as the ONC population (30-60\% if the Trapezium cluster is removed from consideration).

\item Our results suggest that what is normally taken in the literature as the ONC should be seen as a mix of several unrelated populations: 1) the youngest population, including the Trapezium cluster and ongoing star formation in the dense gas inside the nebula, 2) the young foreground population, dominated by the NGC 1980 cluster, and 3) the poorly constrained population of foreground and background Galactic field stars.

\item We re-determine the mean radial velocity for the Trapezium and NGC 1980 clusters to be $25.4\pm3.0$ km/s and $24.4\pm1.5$ km/s respectively, or indistinguishable within the errors, and similar to the radial velocity of the Orion A cloud, suggestive of a genetical connection between the two.

\item We identify a hitherto unrecognized group of about 50 YSOs  West of L1641N (L1654W) that we speculate is either a foreground group ramming into the Orion A cloud, or a slightly older sibling of NGC 1641N, leaving the cloud. 

\item This work supports a scenario where the ONC and L1641N are not directly associated with NGC 1980, i.e., they are not the same population emerging from its parental cloud but are instead distinct overlapping populations. This calls for a revision of most of the observables in the benchmark ONC region (e.g., ages, age spread, mass function, disk frequency, etc.).

\end{itemize}

\acknowledgements We thank the referee, John Bally, for comments that improved the manuscript. We also thank John Tobin, Nicola Da Rio, and Lynne Hillenbrand for comments and clarifications that improved the presentation of results.  H. Bouy is funded by the Ram\'on y Cajal fellowship program number RYC-2009-04497. We acknowledge support from the Faculty of the European Space Astronomy Centre (ESAC). This publication is supported by the Austrian Science Fund (FWF). We thank Calar Alto Observatory for allocation of director's discretionary time to this programme.

Funding for SDSS-III has been provided by the Alfred P. Sloan Foundation, the Participating Institutions, the National Science Foundation, and the U.S. Department of Energy Office of Science. The SDSS-III web site is http://www.sdss3.org/.

SDSS-III is managed by the Astrophysical Research Consortium for the Participating Institutions of the SDSS-III Collaboration including the University of Arizona, the Brazilian Participation Group, Brookhaven National Laboratory, University of Cambridge, University of Florida, the French Participation Group, the German Participation Group, the Instituto de Astrofisica de Canarias, the Michigan State/Notre Dame/JINA Participation Group, Johns Hopkins University, Lawrence Berkeley National Laboratory, Max Planck Institute for Astrophysics, New Mexico State University, New York University, Ohio State University, Pennsylvania State University, University of Portsmouth, Princeton University, the Spanish Participation Group, University of Tokyo, University of Utah, Vanderbilt University, University of Virginia, University of Washington, and Yale University.

This publication makes use of data products from the Wide-field Infrared Survey Explorer, which is a joint project of the University of California, Los Angeles, and the Jet Propulsion Laboratory/California Institute of Technology, funded by the National Aeronautics and Space Administration.

Based on observations obtained with XMM-Newton, an ESA science mission with instruments and contributions directly funded by ESA Member States and NASA.

This research used the facilities of the Canadian Astronomy Data Centre operated by the National Research Council of Canada with the support of the Canadian Space Agency.   

This publication makes use of data products from the Two Micron All Sky Survey, which is a joint project of the University of Massachusetts and the Infrared Processing and Analysis Center/California Institute of Technology, funded by the National Aeronautics and Space Administration and the National Science Foundation.

This work is based in part on observations made with the Spitzer Space Telescope, which is operated by the Jet Propulsion Laboratory, California Institute of Technology under a contract with NASA.

\bibliographystyle{aa} 

\bibliography{/Users/jalves/Dropbox/Shared/Herve/ourion/bib}

\end{document}